 \documentstyle[12pt]{article}
  
\catcode`@=11
\def\secteqno{\@addtoreset{equation}{section}%
\def\theequation{\thesection.\arabic{equation}}}
\catcode`@=12
\secteqno
\newcommand{\be}{\begin{equation}}
\newcommand{\ee}{\end{equation}}
\newcommand{\bea}{\begin{eqnarray}}
\newcommand{\eea}{\end{eqnarray}}
\newcommand{\bref}[1]{(\ref{#1})}
\newcommand{\nn}{\nonumber}

\newcommand{\Tb}{{\overline\theta}}
\newcommand{\bi}{\begin{enumerate}}
\newcommand{\ei}{\end{enumerate}}
\newcommand{\A}{\alpha} \newcommand{\B}{\beta} 
 \newcommand{\D}{\delta} 
\newcommand{\ep}{\epsilon} 
\newcommand{\T}{\theta} 

\newcommand{\lam}{\lambda} 
          
\newcommand{\z}{\zeta}          
\newcommand{\h}{\eta}           
           
\newcommand{\W}{\Omega}

\newcommand{\epb}{{\overline\epsilon}}
\newcommand{\ba}{\overline }
\def\6{\partial}\def\7{\tilde}\def\8{\hat}
\def\bu{{\bf u}}\def\bv{{\bf v}}
\def\pa{\partial}

\def\CC{{\cal C}}\def\CL{{\cal L}}
\def\CT{{\cal T}}
\def\t{\tilde}\def\too{\quad\to\quad}
\def\vs{\vskip 4mm}\def\={{\;=\;}}\def\+{{\;+\;}}
\def\ten{{{\frac{T}2}}}\def\ggt{{\Gamma_{11}}}

\newcommand{\slP}{/ {\hskip-0.25cm{P}}}
\begin{document}
\thispagestyle{empty}
\hfill March 19, 2008

\hfill Toho-CP-0888

\vskip 20mm
\begin{center}
{\Large\bf Massive Rigid String Model and its \\
Supersymmetric Extension}
\vskip 6mm
\medskip

\vskip 10mm
{\large ~Kiyoshi Kamimura and Daisuke Shiseki}

\parskip .15in

{\it
 Department of Physics, Toho University, Funabashi, 274-8510, Japan}\\
 {\small e-mail:\ 
 {kamimura@ph.sci.toho-u.ac.jp} }\
 \

\medskip
\end{center}
\vskip 40mm
\begin{abstract}
We discuss a rigid string model proposed by Casalbuoni and Longhi.
Constraints for the massive states are solved to find the physical states
and the mass spectrum. We also find its supersymmetric extension 
with the kappa symmetry. The supersymmetry transformations are found starting
from on-shell transformations using the Dirac bracket. 

\end{abstract}

\vskip 4mm
\setcounter{page}{1}
\parskip=7pt
\newpage

\section{Introduction}

The superstring theories are promising approaches to the
unified theory of fundamental interactions.
The original bosonic string action is given in a geometrical form of the
Nambu-Goto action \cite{Nambu:1970, Goto:1971ce}. A possible modification 
was considered by adding an extrinsic curvature term\cite{Polyakov:1986cs}.
It is sometimes called the rigid string since the additional term
introduces rigidity of string \cite{Kleinert:1986bk,Curtright:1986vg,Curtright:1986ed}.

  The "rigid string" model we will discuss in this paper is one originally
proposed by Casalbuoni and Longhi\cite{Casalbuoni:1974tq}.
It is a bi-local type model and the action is proportional
to the area of the world sheet swept out by the rigid string.
In contrast to the rigid string of \cite{Polyakov:1986cs} string rigidity is
built in the action geometrically. 
It is described by two end point coordinates $x^\mu_j(\tau), (j=1,2)$ and
the string stretches straight in the relative coordinate
direction $r^\mu=x_1^\mu-x_2^\mu$ then the action can be written as\footnote{
There is a similar approach known as "straight string model" having different
lagrangian \cite{Pronko:1992gc,Kalashnikova:1997jg}.}
\bea
S=-T\;\int\;\left(\frac12\sqrt{|dx_1^{[\mu} r^{\nu]}|^2}+
\frac12\sqrt{|dx_2^{[\mu} r^{\nu]}|^2}\;\right)=\int\;L\;d\tau,
\eea
where $dx_j^{[\mu} r^{\nu]}$ is the surface element spanned by
$dx_j^{\mu}$ and $r^{\nu}$.
Then we have a NG type Lagrangian
\bea
L&=&-\frac{T}2
\left(\sqrt{(\dot x_1r)^2-r^2(\dot x_1)^2}+\sqrt{(\dot x_2r)^2-r^2(
\dot x_2)^2}\right).
\label{LagNG}\eea
It is important that direction of the relative coordinate 
$r^\mu=x_1^\mu-x_2^\mu$
is dynamically specified through the constraints from the action.

In \cite{Casalbuoni:1974tq} the massless sectors of the model was examined in
detail and it was shown that the model describes massless gauge particles,
photon, gravitons,
and so on, with appropriate polarization properties.
The action \bref{LagNG} also allow "massive string states" whose classical
motion is corresponding to a rotating rod. The mass square of the string
is proportional to the angular momentum.

In the first part of this paper we discuss the massive sector of the model
in the hamiltonian formalism. There appear the second class constraints
specifying that the relative coordinate and momentum are orthogonal to
the total momentum.
Thus the internal motion is described by the transverse coordinates and
momenta satisfying further constraints from the lagrangian  \bref{LagNG}.
The internal symmetries are SU(1,1) as well as the SO(d-2) and
the physical states are constructed and their mass spectrum is determined.

In the second part we consider a supersymmetric extension of the 
model.\footnote{The supersymmetic string action with extensic curvature
has beed developed in \cite{Curtright:1987mr}
}
The NG type Lagrangian \bref{LagNG} can have space-time supersymmetry
by introducing target space spinors.
As usual the local fermionic invariance, the kappa symmetry, requires
additional WZ type lagrangian. Although it is constructed 
cohomologically\cite{Henneaux:1984mh,DeAzcarraga:1989vh} in the superstring 
theories we cannot apply it since the space of the coordinates are two world 
lines rather than  two dimensional world sheet. However
we can construct a kappa invariant action in a similar form as the
D0 particle WZ action\cite{Bergshoeff:1996tu}.

The organization of the paper is as follows.
In the next section we make the hamiltonian analysis of the lagrangian
\bref{LagNG} and show the massive and massless sectors as possible
branches of the system.
In sect.III we quantize the bosonic model to find the physical states.
In sect.IV we generalize it to the supersymmetric model and find
possible kappa symmetric extension.
In sect.V we analyze it in the hamiltonian formalism to find the 
supersymmetry generators. 
The last section is devoted to summary and discussions.
In Appendix we give a form of off-shell supersymmetry transformations.

\section{Bosonic Rigid String; Constraints and Hamiltonian}

The bosonic lagrangian of the rigid string \bref{LagNG} is written as
\bea
L&=&-{\ten}\sqrt{r^2}\left(\sqrt{-(v_{1\perp})^2}+\sqrt{-(v_{2\perp})^2}
\right),
\label{Lag2}\eea
where
\be
v_i^\mu=\dot x_i^\mu,\qquad
v_{i\perp}^\mu=(\h^{\mu\nu}-\frac{r^\mu r^\nu}{r^2})v_{i\nu},\qquad
r^\mu=x^\mu_1-x^\mu_2,\qquad (i=1,2)
\ee
and ${T}$ is a constant\footnote{
Models with non-constant $T=T(r^2)$ are discussed in \cite{Fujigaki:1976wn}.}
with dimension $[T]=[{\ell}^{-2}]$.
It is a singular lagrangian and examined by using the generalized hamiltonian
formalism\cite{Dirac:1950pj}.
The momenta conjugate to the coordinates $x_i^\mu, (i=1,2)$ are
\be
p_{i\mu}=\frac{\ten\sqrt{r^2}}{\sqrt{-(v_{i\perp})^2}}\;v_{i\perp\mu}.
\label{defmom}\ee
They are not independent but satisfy following four primary constraints
\be
\phi_i=\frac12\left(p_i^2+(\ten)^2r^2\right)=0,\qquad \psi_i=p_ir=0.
\label{pricon}\ee
The generalized Hamiltonian $p_i\dot x_i-L$ is\cite{Kamimura:1981fe,
Batlle:1985ss}
\be
H=\frac{\sqrt{-(v_{i\perp})^2}}{T\sqrt{r^2}}\;
\left(p_{i}^2+(\ten)^2{r^2}\right)+\frac{v_ir}{r^2}\;p_ir-
\frac{\sqrt{-(v_{i\perp})^2}}
{T\sqrt{r^2}}
\left(p_{i\mu}-\frac{{\ten}\sqrt{r^2}}{\sqrt{-(v_{i\perp})^2}}v_{i\perp\mu}
\right)^2.
\label{Ham0}\ee
The last term is "squire of the definition of momenta" \bref{defmom}
and does not contribute in the equations of motion.
The hamiltonian is expressed as a sum of the four primary constraints,
\bea
H&=&\lam_j\;\phi_{j}+\mu_j\;\psi_j=
\lam_j\;\frac12\left(p_{j}^2+(\ten)^2{r^2}\right)+\mu_j\;(p_jr),
\label{HamD}\eea
where the multipliers $\lam_j$ and $\mu_j$  are given in terms
of undetermined velocities as in \bref{Ham0},
\bea
\lam_j=\frac{\sqrt{-(v_{j\perp})^2}}{{\ten}\sqrt{r^2}},\qquad
\mu_j=\frac{v_jr}{r^2} .
\label{multiplilammu}\eea
Note both $\lam_1$ and  $\lam_2$ are positive in order that
the Euler-Lagrange equations are reproduced from the Hamilton equations.

The consistency condition that the primary constraints are conserved in time
gives
\bea
\pa_\tau\phi_i&=&\pmatrix{2(\ten)^2r^2&-p_1p_2-(\ten)^2r^2\cr p_1p_2+(\ten)^2
r^2&-2(\ten)^2r^2}\pmatrix{\mu_1\cr \mu_2}=0,
\\
\pa_\tau\psi_i&=&\pmatrix{-2(\ten)^2r^2&-p_1p_2-(\ten)^2r^2\cr p_1p_2+(\ten)^2
r^2&2(\ten)^2r^2}\pmatrix{\lam_1\cr \lam_2}=0.
\eea
To have non trivial motion it is necessary that the determinant of the
above matrices vanish. It is satisfied in the following two cases
\bea
p_1p_2+(\ten)^2r^2=\pm 2(\ten)^2r^2.
\label{seccon}\eea
The upper sign solution of \bref{seccon} gives
\be
p_1p_2-(\ten)^2r^2=0,\quad \mu_1=\mu_2,\quad \lam_1=-\lam_2.
\ee
Since $\lam_1$ and $\lam_2$ have opposite signs this case is discarded.
The lower sign solution of \bref{seccon} gives
\be
\chi\equiv p_1p_2+3(\ten)^2r^2=0,\quad \mu_1=-\mu_2,\quad \lam_1=\lam_2\equiv
\lam.
\label{case10}\ee
The $\chi=0$ is the secondary constraint and it further requires
\be
\pa_\tau\chi=(\mu_1-\mu_2)\,T^2\,r^2=2\,\mu_1\,T^2\,r^2=0.
\ee
It gives either $\mu_1=0$ or $r^2=0$. The former case gives
\be
 \mu_1=-\mu_2=0
\label{case1}\ee
and no more constraint appears. It gives massive states as we will
study in below.
In the latter case $r^2=0$ is the tertiary constraint and
the set of constraints is
\be
p_1^2=p_2^2=p_1p_2=p_1r=p_2r=r^2=0.
\label{case2}\ee
They are first class constraints and conserved for any $\lam_i, \mu_i$.
The constraints \bref{case2} mean the massless states $P^2=(p_1+p_2)^2=0$.
Thus the lagrangian system is describing both massless and massive
sectors. The massless sectors are examined in detain in 
\cite{Casalbuoni:1974tq} and it was shown that massless gauge particle 
states appear in the quantized spectrum.
In the following we will discuss the massive sector.
\vs

We consider the case of \bref{case10} and  \bref{case1} then the hamiltonian
becomes
\bea
H&=&\lam(\phi_{1}+\phi_{2})=
\frac{\lam}{2}(p_{1}^2+p_{2}^2+\frac{T^2}2{r^2})\equiv\;\lam\;\phi.
\label{HamD2}\eea
Introducing center of mass and relative coordinates as
\be
P_\mu=p_{1\mu}+p_{2\mu},\quad X^\mu=\frac12(x^\mu_1+x^\mu_2),\quad
q_\mu=\frac12(p_{1\mu}-p_{2\mu}),\quad r^\mu=x^\mu_1-x^\mu_2,
\ee
the primary constraints \bref{pricon} and secondary constraints \bref{seccon}
are
\be
Pr=Pq=qr=q^2-(\ten)^2r^2=0,\quad {\rm and} \quad
\phi=\frac14P^2+(q^2+(\ten)^2r^2)=0.
\label{allconst}\ee
$\phi=0$ is the first class constraint and appearing in the Hamiltonian with
arbitrary multiplier $\lam(\tau)$. Other 4 constraints are the second class.
We first eliminate $Pr=Pq=0$ at classical stage using a canonical
transformation\cite{Kamimura:1977dv,Dominici:1978yu}.
We also consider the bosonic system in 4
dimensions so that the symmetry algebra in the transverse space is SO(3).
It leaves transverse relative three coordinates ${\bf u}$ and
relative three momenta ${\bf v}$ subject to the constraints
\be
\CT_1 \equiv \frac1{2T}(\bv^2-(\ten)^2\bu^2)=0,\quad
\CT_2 \equiv \frac12\bv\bu=0
\label{T1T2}
\ee
and the first class constraint
\be
\phi=\frac14P^2+{2T}\;\CT_0=0,\qquad
\CT_0 \equiv\frac1{2T}(\bv^2+(\ten)^2\bu^2).
\label{phiT3}\ee
The last one fixes the mass of the system as
\be
M^2=-P^2={8\,T}\; \CT_0.
\ee
Classically the mass is determined in terms of the angular momentum
\be
{\CL}= {\bf u}\times {\bf v}
\label{SO3L}\ee
as follows. Using constraints $\CT_1=\CT_2=0$ in \bref{T1T2},
\be
{\CL}^2= ({\bf u}\times {\bf v})^2=({\bf u})^2({\bf v})^2-({\bf u}{\bf v})^2=
(\ten)^2({\bf u}^2)^2.
\ee
Then the (mass$)^2$ is proportional to the length of the angular momentum,
\be
M^2=-P^2=2{T}^2({\bf u}^2)=4{T}\;|{\cal L}|.
\label{m2classical}\ee

\section{Quantization and Physical States}

We discuss the quantization of the transverse variables $(\bu,\bv)$
subject to the constraints \bref{T1T2}. The $\CT_0$ , therefore the
Hamiltonian, is diagonalized using ladder operators
\be
a_r=\frac{1}{\sqrt{T}}(v_r-i{{\ten}}u_r),\qquad
a_r^\dagger=\frac{1}{\sqrt{T}}(v_r+i{{\ten}}u_r),\qquad
[a_r, a_s^\dagger]=\D_{rs}.
\ee
The operators $ \CT_1$ and $\CT_2$ in \bref{T1T2} and $ \CT_0$
in \bref{phiT3} are
\be
\CT_-=\CT_1-i\;\CT_2=\frac12a_r^2,\qquad
\CT_+=\CT_1+i\;\CT_2=\frac12(a_r^\dagger)^2,\qquad
\CT_0 =\frac12(a_r^\dagger a_r+\frac32)
\label{phiT32}\ee
and form  the SU(1,1) algebra 
\be
[\CT_0,\CT_\pm]=\pm\CT_\pm,\qquad [\CT_+,\CT_-]=2\CT_0.
\ee
The first class constraint $\phi=0$ in \bref{phiT3} is hermitic and is
imposed as the physical state condition
\be
\phi\;|phys\rangle=\left(\frac14P^2+2T\;\CT_0+c_0\right)|phys\rangle=0,
\ee
where the constant $c_0$ is an ordering ambiguity.
Therefore the mass of the physical state $|phys\rangle$ is determined as
\be
M^2\;|phys\rangle=-P^2\;|phys\rangle=4
\left(2T\;\CT_0+c_0\right)|phys\rangle.
\ee
The second class constraints $\CT_1=\CT_2=0$ in \bref{T1T2}
at quantum theory are imposed as a physical
state condition {\it a.la.} Gupta-Bleuler,
\be
\CT_-\;|phys{\rangle}=\frac12\;a_r^2\;|phys{\rangle}=0.
\label{PSC}
\ee
We can solve them to find the physical states and their mass spectrum.

In order to find the physical states we consider eigenstates of the
angular momentum ${\CL}$ in \bref{SO3L},
\be
\CL_r=-i\;\ep_{rst}\;a_s^\dagger a_t.
\label{SO3aa}\ee
The internal rotations SO(3) is a symmetry of the model and the generators
${\CL_r}$ commute with those of SU(1,1), $\CT_\pm$ and $\CT_0$.
$\CL_3=-i(a_1^\dagger a_2-a_2^\dagger a_1)$ is not diagonal in $a_r$
but can be diagonalized by a unitary transformation
\be
{\bf a}=U {\bf b},\qquad {\bf a}^\dagger ={\bf b}^\dagger U^\dagger,\qquad
U=\pmatrix{-1/\sqrt2 &0& 1/\sqrt2 \cr -i/\sqrt2 &0& -i/\sqrt2 \cr 0&1&0}.
\ee
In terms of $b_r$ the  SO(3)  generators \bref{SO3aa} are
\bea
\CL_3&=&{b}^\dagger_1b_1-{b}^\dagger_3b_3, \qquad
\nn\\
\CL_+&=&\CL_1+ i \CL_2=\sqrt2\;(\;{b}^\dagger_1b_2+\;{b}^\dagger_2b_3\;),
\qquad\nn\\
\CL_-&=&\CL_1- i \CL_2=\sqrt2\;(\;{b}^\dagger_2b_1+\;{b}^\dagger_3b_2\;).
\eea
The SU(1,1)  generators in terms of  $b_r$   are
\bea
\CT_0&=&\frac12({b}^\dagger_1b_1+{b}^\dagger_2b_2+{b}^\dagger_3b_3+\frac32)
, \qquad
\nn\\
\CT_+&=&\CT_1+ i \CT_2=\;\frac12({b^\dagger_2}^2-2\;{b}^\dagger_1b^\dagger_3
\;), \qquad
\nn\\
\CT_-&=&\CT_1- i \CT_2=\;\frac12({b_2}^2-2\;{b}_1b_3)\;.
\eea

General Fock states are constructed by applying $b^\dagger_r$'s
on the ground state $|0{\rangle}$, 
\be
\frac1{\sqrt{j_1!j_2!j_3!}}
( b^\dagger_1)^{j_1}(b^\dagger_2)^{j_2}(b^\dagger_3)^{j_3}|0{\rangle},\qquad
b_r|0{\rangle}=0.
\ee
They are the eigenstates of $\CT_0$ with eigenvalue $\frac12(j_1+j_2+j_3+
\frac32)\equiv\frac12( j+\frac32)$. The number of such states for a fixed 
value of $j$ is
\be
\sum_{j_1=0}^j \sum_{j_2=0}^{j-j_1}\,1=\frac{(j+1)(j+2)}{2}.
\ee
They are decomposed into irreducible representations of SO(3),
\be
\frac{(j+1)(j+2)}{2}=\sum_{n=j,j-2,j-4,..}(2n+1).
\ee
That is sum of spin $j, j-2, j-4,...$ multiplets.
Among them the highest spin $j$ multiplet is constructed from
\be
|j,-j{\rangle}=\frac1{\sqrt{j!}}(b^\dagger_3)^{j}|0{\rangle},\quad
\label{jmj}\ee
by multiplying $\CL_+$ successively.
Here $|j,m{\rangle}$ is the eigenstate of $\CL^2$ and $\CL_3$ with eigenvalues
$j(j+1)$ and $m, (|m|\leq j)$.
They are satisfying the physical state condition \bref{PSC}
since $\CT_-|j,-j{\rangle}=0$ and $[\CT_-,\CL_+]=0$. Other low lying states
with spin $j-2, j-4,...$ are unphysical states and are given by
\bea
|j-2r,m{\rangle}=N\;(\CT_+)^r(\CL_+)^{m+j-2r}(b_3^\dagger)^{j-2r}|0{\rangle},
\quad
r=1,2,..,[\frac{j}2], \; |m|\leq j-2r.
\eea
with a normalization factor $N$.

\vs
In summary the physical states are
\be
|j,m{\rangle}=\sqrt{ \frac{(j-m)!}{(2j)!(j+m)!}}\;(\CL_+)^{m+j}
\frac1{\sqrt{j!}}(b^\dagger_3)^{j}|0{\rangle},\qquad
0\leq j,\;\; -j\leq m\leq j,
\ee
and the mass of the states is
\be
M^2|j,m{\rangle}=4(2T\;\CT_0+c_0)\;|j,m{\rangle}=(4T\;(j+\frac32)+4c_0)\;
|j,m{\rangle}.
\label{Qspect}
\ee
where $c_0$ coming from the ordering ambiguity is not determined,
for example from the Lorentz invariance.
The spectrum \bref{Qspect} is corresponding to the classical one in 
\bref{m2classical} that the mass is coming from the internal rotation energy.
They have maximal spin lying on the leading Regge trajectory
and their motion is corresponding to rotating rod classically.

In the above we have eliminated two of the second class constraints $Pr=Pq=0$
classically. Alternatively we could impose them in quantum theory using 
covariant
oscillators $a_\mu=\frac{1}{\sqrt{T}}(q_\mu-i{{\ten}}r_\mu)$.
In terms of them five constraints in \bref{allconst} are
\bea
&&L_{-2}{\;\equiv\;}\frac12(a_\mu^\dagger)^2,\qquad
L_{-1}{\;\equiv\;}\frac{1}{\sqrt{2T}}\;Pa^\dagger,
\qquad L_0 {\;\equiv\;}\frac{P^2}{4T}+(a_\mu^\dagger a^\mu)
\nn\\
&&L_{2}{\;\equiv\;}\frac12(a_\mu)^2,\qquad
L_1{\;\equiv\;}\frac{1}{\sqrt{2T}}\;Pa,\qquad
\label{phiT323}\eea
They are obtained from the Virasoro generators of the NG string by truncating
the
higher oscillator modes; $a_{n\mu}=0,(n\geq2)$. In contrast to the  Virasoro
generators, which are the first class set classically, the constraints
\bref{phiT323} are not first class. However we can impose them at quantum
theory as the physical state conditions,
\be
L_{2}|phys{\rangle}=L_{1}|phys{\rangle}=(L_{0}-\A_0)|phys{\rangle}=0.
\ee
They are solved as above by using the Lorentz transformation from the rest
frame.

\section{Supersymmetric Model}
The Nambu-Goto action for the superstring is
\be
L^{NG}=-\frac{T}2\left(\sqrt{(v_1r)^2-r^2(v_1)^2}+\sqrt{(v_2r)^2-r^2(v_2)^2}
\right),
\label{LagNGsuper}
\ee
where $v_i$'s are super invariant velocities.
First we leave them in a general form as
\be
v_i^\mu=\dot x_i^\mu+iB_i^{jk}\Tb_j\Gamma^\mu\dot\T_k.
\label{susyv}
\ee
In this section we consider Majorana-Weyl spinors  $\T_j, (j=1,2)$
in 10 dimensions and
$(\CC\Gamma^A), (A=0,1,2,...,9, 11)$ have symmetric gamma 
indices.\footnote{We use the mostly positive metric 
$\h_{\mu\nu}=(-;+...+)$ and the Clifford algebra is
$\{\Gamma_\mu,\Gamma_\nu\}=2\h_{\mu\nu}.$}
The relative coordinate is
\be
r^\mu=x_1^\mu-x_2^\mu
\ee
since additional fermionic contributions, if any, can be absorbed into
$x_j^\mu$ by redefinition.

The global susy transformation is determined from the susy  invariance of
$v_i$,
\be
\D_\ep\T_i=\ep_i,\quad \D_\ep x_i^\mu=-iB_i^{jk}\epb_j\Gamma^\mu\T_k,\qquad
\too \D_\ep v_i^\mu=0.
\label{supert}\ee
The susy transformation of $r^\mu$ is
\bea
\D_\ep r^\mu&=& \D (x_1^\mu-x_2^\mu)=-i(B_1^{jk}-B_2^{jk})\epb_j\Gamma^\mu\T_k.
\eea
For the invariance of  $r^\mu$ it is sufficient to choose the
coefficients $B_i^{jk}$ as
\bea
B_1^{jk}=B_2^{jk}\equiv B^{jk},
\label{Bijk}\eea
where  $B^{jk}$ is taken to be symmetric matrix since the anti-symmetric part
can be absorbed into the definition of $x_i$ in this case.

Kappa transformation is the local fermionic symmetry under which
\be
\D_\kappa\T_j=\kappa_j(\tau),\qquad \D_\kappa
 x_i^\mu= -iB^{jk}\Tb_j\Gamma^\mu\kappa_k(\tau),
\ee
then
\bea
\D_\kappa v_i^\mu&=&
-2i\;B^{jk}\;\dot\Tb_j\Gamma^\mu\kappa_k.
\eea
Under the choice \bref{Bijk} $ r^\mu$ is kappa invariant also,
\bea
\D_\kappa r^\mu&=& \D (x_1^\mu-x_2^\mu)=0.
\eea

The $L^{NG}$ is invariant under the super transformation \bref{supert} with
\bref{Bijk} while it transforms
under the kappa transformation as
\be
\D_\kappa L^{NG}=p^i\D v_i
= -2i\;B^{jk}\;\dot\Tb_j\slP\kappa_k,
\ee
where $p^i_{\mu}=\frac{\pa L^{NG}}{\pa v_i^\mu}$.
To compensate it we consider a possible additional lagrangian,
corresponding to WZ one. Usually it appears
from the discussions of non-trivial Chevalley-Eilenberg cohomology of the 
super Poincare group\cite{Henneaux:1984mh,DeAzcarraga:1989vh}.
In the case of superstring, there exists a closed susy invariant three form
$\W^3=d\W^2$. The $\W^2$ is not an element of the
Chevalley-Eilenberg cohomology and works as the WZ term.
In the present case the world sheet is degenerated to the world lines
at the boundaries due to the bi-local nature. 
However we can find the additional action in a form similar to the D0 particle
WZ action\cite{Bergshoeff:1996tu} by replacing its mass to 
$r\equiv\sqrt{r^2}$,
\bea
L^{WZ}&=&i\,r\,C^{jk}\Tb_j\ggt\dot\T_k,
\label{LagWZ}\eea
where $C^{jk}$ is some constant matrix.
It transforms under the susy
\bea
\D_\ep L^{WZ}&=&i\,r\,C^{jk}\epb_j\ggt\dot\T_k=\pa_\tau(i\,r\,C^{jk}\epb_j\ggt
\T_k)-i\,\dot r\,C^{jk}\epb_j\ggt\T_k.
\label{delLWZep}\eea
It is (pseudo) invariant if the supersymmetry parameters $\ep_J$ is
restricted by
\be \ep_j\;C^{jk}=0. \label{susycond}\ee
However since $\dot r=0$ is a result of the equation of motion,
as we will see in \bref{case1s},
\be
\mu_1-\mu_2=\frac{(r\dot r)}{r^2}=0
\label{case11}\ee
$ L^{WZ}$ is invariant for any  supersymmetry parameters $\ep_J$
{\it on-shell} and the super charges will be introduced as in 
\bref{supercharge}.\footnote{
If a variation of a lagrangian $ \delta L= (\frac{\partial
L}{\partial q}-\frac d {dt} \frac{\partial L}{\partial\dot q})\delta
q+\frac d {dt} (\frac{\partial L}{\partial\dot q}\delta q) 
$ is written non-trivially in a form $\frac{d}{dt}F+(eom)$, then 
$G=(\frac{\partial L}{\partial\dot q}\delta q-F)$ is a conserved quantity, 
where $(eom)=0$ using equations of motion. }

The $L^{WZ}$ transforms under the kappa
\bea
\D_\kappa L^{WZ}&=&
-2ir\,C_+^{kj}\;\dot\Tb_j{\ggt}\kappa_k
-i\,\dot r\,C^{jk}\Tb_j{\ggt}\kappa_k +
\pa_\tau(i\,r\,C^{jk}\Tb_j{\ggt}\kappa_k ),
\eea
where $C^{jk}_\pm=\frac12(C^{jk}\pm C^{kj})$ are symmetric and anti-symmetric
parts of the constant matrix $C^{jk}$.
The kappa variation of the total lagrangian is
\bea
\D_\kappa L^{tot}&=&
-2i\;\dot\Tb_j(B^{jk}\,\slP+C^{jk}_+\,r\,{\ggt})\kappa_k
-i\,\dot r\,C^{jk}\;\Tb_j{\ggt}\kappa_k
+\pa_\tau(ir\,C^{jk}\Tb_j{\ggt}\kappa_k ).
\eea
The  action is kappa invariant if the kappa functions $\kappa_k(\tau)$'s
satisfy
\be
C^{jk}\,\kappa_k=0
 \label{kapcon1L}\ee
and
\be
\left(\D^{jk}  \,\slP+(B^{-1}C_+)^{jk}\,r\,{\ggt}\right) \kappa_k =
\left(\D^{jk}  \,\slP-(B^{-1}C_-)^{jk}\,r\,{\ggt}\right) \kappa_k =0.
 \label{kapcon2L}\ee
In order to have non-trivial kappa transformations it is necessary to hold
\be
[(B^{-1}C_-),C]_-\sim C.
\ee
There exists such matrices, for example 
\bea
B&=&\frac{1}{\B}\pmatrix{1-\B&1\cr1&1+\B},\qquad
C=k\pmatrix{\B+1&\B+1\cr\B-1&\B-1},\qquad
\label{BBCC}\\
C_-&=&k\pmatrix{0&1\cr-1&0},\qquad (B^{-1}C_-)^2=
(\frac{k}{\B})^2\pmatrix{-1&-(1+\B)\cr(1-\B)&1}^2=
k^2,
\label{BBCC2}\eea
where $k$ and $\B$ are non-zero real constants.
Note if we take $k=\sqrt{2}\,{T}$,
\be
\left(\slP-(B^{-1}C_-)\,r\,{\ggt}\right)^2=
P^2+2\,T^2\,r^2=0
\ee
using constraints from the lagrangian as we will see shortly. 
In this case we can write
\bea
\left(\slP-(B^{-1}_+C_-)\,r\,{\ggt}\right)=\frac{1-\Gamma_\kappa}2 \;2\slP, 
\qquad 
\Gamma_\kappa\equiv\sqrt{2}\,T\,r\,\ggt\frac{1}{\slP},\qquad \Gamma_\kappa^2=1
\label{BPScon3}\eea
and $\frac{1\pm\Gamma_\kappa}2 $ work as the
projection operators for the kappa transformations.

\vs
\section{Hamiltonian Formalism of Supersymmetric Model}

We are going to discuss the total Lagrangian \bref{LagNGsuper}+\bref{LagWZ},
\be
L^{tot}=-\frac{T}2\left(\sqrt{(v_1r)^2-r^2(v_1)^2}+\sqrt{(v_2r)^2-r^2(v_2)^2}
\right)
+i\,r\,C^{jk}\Tb_j\ggt\dot\T_k.
\label{Lagtotsuper}
\ee
with $v_i^\mu=\dot x_i^\mu+iB^{jk}\Tb_j\Gamma^\mu\dot\T_k.
$
Since $L^{WZ}$ does not depends on $\dot x_j$ the bosonic primary constraints
appear in the same forms as in the bosonic model, \bref{pricon},
\be
\phi_i=\frac12(p_i^2+(\ten)^2r^2)=0,\qquad \psi_i=p_ir=0.
\label{priconsp}\ee
We define the fermionic momentum $\pi^k$  conjugate to $\T_k$
by the {\it right derivative}
\bea
\pi^k&=&\frac{\pa^rL^{tot}}{\pa\dot\T_k}=
iB^{jk}\Tb_j\slP+i\,r\,C^{jk}\Tb_j\ggt,
\eea
then the fermionic primary constraints are
\bea
\z^k&=&\pi^k-i\Tb_j(B^{jk}\slP+C^{jk}\,r\,\ggt)=0.
\eea
The hamiltonian is expressed as a sum of the primary constraints,
\bea
H&=&\lam_j\;\phi_{j}+\mu_j\;\psi_j+\z^k\;\rho_k.
\label{HamDsp}\eea
where $\lam_j$ and $\mu_j$  are the bosonic multipliers given in the
same forms as in \bref{multiplilammu} but with the $v_i^\mu$ in \bref{susyv}
and $\rho_k=\dot\T_k$ is the fermionic multiplier.
The consistency condition that the primary constraints are conserved in time
gives
\bea
\pa_\tau\phi_i&=&\pmatrix{2(\ten)^2r^2&-p_1p_2-(\ten)^2r^2\cr p_1p_2+(\ten)^2
r^2&-2(\ten)^2r^2}\pmatrix{\mu_1\cr \mu_2}=0,
\label{cons1}\\
\pa_\tau\psi_i&=&\pmatrix{-2(\ten)^2r^2&-p_1p_2-(\ten)^2r^2\cr p_1p_2+(\ten)^2
r^2&2(\ten)^2r^2}\pmatrix{\lam_1\cr \lam_2}+i\,r\,C^{jk}\Tb_j\ggt\rho_k
\pmatrix{1\cr-1}=0.
\nn\\ \label{cons2}\\
\pa_\tau\z^k&=&-2 i\ba\rho_j(B_+^{jk}\slP+C_+^{jk}\,r\,\ggt)
-i\Tb_jC^{jk}\ggt\,r\,(\mu^1-\mu^2)=0.
\label{cons3}\eea
As in the bosonic case we get the secondary constraint from \bref{cons1},
\be
\chi\equiv p_1p_2+3(\ten)^2r^2=0,\qquad \mu_1=-\mu_2.
\label{case10s}\ee
Using it \bref{cons2} requires
\be
\lambda_1=\lambda_2,
\ee
and 
\be C^{jk}\rho_k=0,
\label{kapcon1}\ee
which for the choice of $C^{jk}$ in \bref{BBCC} restricts $\rho_k$ as
\be
\rho_1=-\rho_2\equiv\rho.
\label{kapcon1a}\ee The secondary constraint $\chi=0$ in \bref{case10s} 
further requires
\be
\pa_\tau\chi=2(\mu_1-\mu_2)\,T^2\,r^2=4\,\mu_1^2\,r^2=0.
\ee
It gives, corresponding to the massive bosonic sector,
\be
 \mu_1=-\mu_2=0
\label{case1s}\ee
and no more constraint appears. 
Finally  \bref{cons3} requires
\be
(B_+^{jk}\slP+C_+^{jk}\,r\,\ggt)\rho_k=0,\quad\to\quad
(\slP-k\,r\,\ggt)\rho=0.
\label{kapcon2}\ee
The conditions \bref{kapcon1} and \bref{kapcon2} are corresponding to
 \bref{kapcon1L} and \bref{kapcon2L}.
 To exist non-trivial $\rho$ the $(\slP-k\,r\,\ggt)$ must be
a projection operator. It determines $k=\sqrt{2}\,T$ as
\be
 (\slP-k\,r\,\ggt)^2=P^2+k^2\;r^2=(p_1+p_2)^2+2\,T^2\,r^2=0.
\ee
Then the Hamiltonian becomes
\bea
H&=&\lam\;(\phi_{1}+\phi_2)+(\z^1-\z^2)\;(\slP-\sqrt{2}\,T\,r\,\ggt)\t\rho,
\label{HamDsp2}\eea
where $\t\rho$ is arbitrary MW spinor but only half components are independent
due to the projector $(\slP-\sqrt{2}\,T\,r\,\ggt)$.
The constraints appearing here belong to the first class.
\bea
\phi_{1}+\phi_{2}=
\frac{1}{2}(p_{1}^2+p_{2}^2+\frac{T^2}2{r^2})\equiv\;\phi
\label{HamD2sp}\eea
generates the $\tau$ reparametrization and determines the mass of the system.
\bea
\z&\equiv&(\z^1-\z^2)\;(\slP-\sqrt{2}\,T\,r\,\ggt)
=\left\{\pi^1-\pi^2+i(\Tb_1+\Tb_2)\slP\right\}(\slP-\sqrt{2}\,T\,r\,\ggt)
\nn\\ \eea
generates the kappa symmetry.
\vs

We have seen the lagrangian is pseudo invariant only {\it on-shell} in
\bref{delLWZep}. Although it is on-shell invariant we can introduce
the global supercharges (see footnote 3),  
\be
Q^j=\pi^j+i\ba\T_k(B^{jk}\slP+C^{jk}\;r\;\ggt).
\label{supercharge}
\ee
 They are conserved and satisfying the super Poincare algebra,
especially \be
\{Q^j,Q^k\}_+=-2\left(B^{jk}(\CC\slP)+C_+^{jk}\;r\;(\CC\ggt)\right).
\ee Here the last term is the central charge, in which
$r=\sqrt{r^2}$ commutes with super Poincare generators,
$(P_\mu,M_{\mu\nu},Q^j)$. 

There is one constraint $qr=(\psi_1-\psi_2)/2=0$ having non-zero Poisson 
brackets with $Q^j$ while other all constraints have weakly zero Poisson 
brackets. 
It does not break the symmetry of the model however.
To clarify it we introduce the Dirac bracket\cite{Dirac:1950pj}, or 
equivalently "stared quantities"\cite{Hanson:1976cn}
which have weakly zero Poisson brackets with all the second class constraints. 
The modified super  Poincare generators 
$({P^*}_\mu,{M^*}_{\mu\nu},{Q^*}^j)=({P}_\mu,{M}_{\mu\nu},{Q^*}^j)$ 
are also conserved and verifying the same  super Poincare algebra. 
The modified supersymmetry transformations, generated by ${Q^*}^j$,
\be
{Q^*}^j=Q^j-\{Q^j,qr\}\frac{-1}{2(q^2+r^2)}(q^2-r^2),
\label{Qstar}\ee
are different from the original ones \bref{supert} and \bref{Bijk}. 
Note there is ambiguity of higher power terms of constraints in the 
Hamiltonian  supercharges in \bref{Qstar}. In Appendix we show 
that there are corresponding supersymmetry transformations of the 
lagrangian \bref{Lagtotsuper} under which it is pseudo invariant 
{\it off-shell}.

\section{Summary and Discussions}

In this paper we have examined the rigid string model proposed
in \cite{Casalbuoni:1974tq}. Especially we have discussed the
massive sector of the model that have not been examined in
\cite{Casalbuoni:1974tq}.
It is quantized and the physical states are constructed explicitly
using the representations of SO(3) and SU(1,1).
The mass of the physical states are determined by their angular momenta and
they are corresponding to the rotating rigid rods. The physical states
are those with the highest angular momentum then lying on the
leading trajectories. States with lower angular momenta, lying on
daughter trajectories are unphysical.
In a similar approach of the straight string model\cite{Pronko:1992gc,
Kalashnikova:1997jg}
a different form of lagrangian was proposed starting from the
NG type. The model is characterized by three first class constraints
corresponding to three local gauge symmetries of the action. We can show that
it is reduced to the present model when we impose
two gauge fixing conditions  appropriately .

We have also examined a possible supersymmetric model with the
kappa symmetry. The $L^{NG}$ action have target space supersymmetries but
it requires an additional $L^{WZ}$ action for the kappa symmetry.
However the number of the independent kappa transformation is a half of the
usual superstring cases. The kappa transformation parameters are restricted
by two projection conditions \bref{kapcon1} and \bref{kapcon2}.

In this model we can examine the BPS states correspondingly.
The bosonic solution with $\T_j=0$ remains under combined super and
kappa transformations if
\bea
\D\T_1&=&\ep_1+\frac{1-\Gamma_\kappa}{2}\;2\slP\t\kappa=0,\qquad
\D\T_2=\ep_2-\frac{1-\Gamma_\kappa}{2}\;2\slP\t\kappa=0,\qquad
\label{BPScon}\eea
where $\frac{1-\Gamma_\kappa}{2}$ is the projection operators for the kappa 
transformations in \bref{BPScon3}. It gives BPS conditions that 
$P_\mu$ and $r$ do not depend on $\tau$. 
Since only a half of $\t\kappa$ is independent there remains 1/4 of the
supersymmetry that preserves the BPS solutions in this model.

In section 4 we have started to construct the supersymmetric model by 
expecting that $v_j^\mu$ and $r^\mu$ are susy invariant and  obtained 
only transformations invariant {\it on-shell}.
We have shown that the {\it off-shell} invariant supersymmetry
transformations are obtained by using the Dirac bracket or the modified
supercharge ${Q^*}^j$. In the corresponding lagrangian transformations
neither $v_j^\mu$ nor $r^\mu$ are susy invariant.
The forms of the  {\it off-shell} transformations are not
simple it is interesting to give any geometrical interpretation
for example in superspace. 

\vs

{\bf Acknowledgements} 

The authors would thank Roberto Casalbuoni and Joaquim Gomis for
useful discussions and encouragements.

\appendix

\section{Appendix: Invariant Susy Transformations}

We will show a sum of following four transformations 1$\sim$4 of the lagrangian
\bref{Lagtotsuper} becomes a total derivative then it is a symmetry 
transformation {\it off-shell}. 
Here we use the lagrangian variables 
\be
p^i_\mu \equiv p^i_\mu(x,v) =\frac{\pa L}{\pa \dot x_i^\mu},\qquad q_\mu=\frac12(p^1_\mu-p^2_\mu)
\ee
and they are satisfying the primary constraints \bref{priconsp}
{\it identically},
\be
\frac12(p_i^2+(\ten)^2r^2)\equiv0,\qquad p_ir\equiv0.
\label{priconspA}\ee
On the other hand the secondary constraint 
\be
\chi= p_1p_2+3(\ten)^2r^2 =  -2(q^2-(\ten)^2r^2) \equiv  -2\,\8\chi
\ee
does not vanish  identically by $p^i_\mu \equiv\frac{\pa L}{\pa\dot x_i^\mu}$.

1: Original supersymmetry transformations  $Q^j\ep_j$, 
\bea
{\D_1}\T_j&=&\ep_j,\quad {\D_1} x_i^\mu=-i\ba\ep_j B^{jk}\Gamma^\mu\T_k,
\quad
\to\quad{\D_1} r^\mu=0,\quad {\D_1} v_i^\mu=0,
\\
{\D_1} L^{tot}&=&{\D_1} L^{WZ}=\pa_\tau(i\,r\,C^{jk}\epb_j\ggt\T_k)
-i\,\dot r\,C^{jk}\epb_j\ggt\T_k.
\label{del1L}\eea

2: Transformation  $F\8\chi$,  
\bea
{\D_2}\T^j&=&0,\quad {\D_2} x_1^\mu=Fq^\mu,\quad 
{\D_2} x_2^\mu=-Fq^\mu,\quad
\\&\to&\quad{\D_2} r^\mu=2Fq^\mu,\quad 
{\D_2} v_1^\mu=\pa_\tau(Fq^\mu),\quad 
{\D_2} v_2^\mu=\pa_\tau(-Fq^\mu),
\nn\\
{\D_2} L^{tot}&=&p^1_\mu\pa_\tau(Fq^\mu)+p^2_\mu\pa_\tau(-Fq^\mu)
-(\mu_1p^1_\mu+\mu_2p^2_\mu)(2Fq^\mu)
\nn\\&=&
\pa_\tau(F(q^2+(\ten)^2r^2))+\dot F{\8\chi}-2F\frac{\dot r}{r}{\8\chi}-4F(\ten)^2
r{\dot r}
\label{del2L}\eea

3:  Transformation  $G\phi$,  
\bea
{\D_3}\T^j&=&0,\quad {\D_3} x_1^\mu=Gp_1^\mu,\quad 
{\D_3} x_2^\mu=Gp_2^\mu,\quad
\\&\to&\quad{\D_3} r^\mu=2Gq^\mu,\quad 
{\D_3} v_1^\mu=\pa_\tau(Gp_1^\mu),\quad 
{\D_3} v_2^\mu=\pa_\tau(Gp_2^\mu),
\nn\\
{\D_3} L^{tot}&=&p^1_\mu\pa_\tau(Gp_1^\mu)+p^2_\mu\pa_\tau(Gp_2^\mu)
-(\mu_1p^1_\mu+\mu_2p^2_\mu)(2Gq^\mu)
\nn\\&=&
\pa_\tau(-2G(\ten)^2r^2)-2G\frac{\dot r}{r}{\8\chi}.
\label{del3L}\eea

4:  Transformation  $\z^j{\8\rho_j}$,  
\bea
{\D_4}\T_j&=&{\8\rho_j},\quad {\D_4} x_i^\mu=-iB^{jk}\Tb_j\Gamma^\mu\8\rho_k,
\quad\to\quad{\D_4} r^\mu=0,\quad 
{\D_4} v_i^\mu=2i\ba{\8\rho}_jB^{jk}\Gamma^\mu\dot\T_k, 
\nn\\
\\
{\D_4} L^{tot}&=&
-2i\;\dot\Tb_j(B^{jk}\,\slP+C^{jk}_+\,r\,{\ggt})\8\rho_k
-i\,\dot r\,C^{jk}\;\Tb_j{\ggt}\8\rho_k
+\pa_\tau(ir\,C^{jk}\Tb_j{\ggt}\8\rho_k ).
\nn\\
\label{del4L}\eea

First we choose $F$ so as to the last terms of \bref{del1L}
and \bref{del2L} cancel,
\be
F=\frac{-1}{4(\ten)^2r}(i\,\epb_jC^{jk}\ggt\T_k),\quad\to\quad
\dot F=\frac{-1}{4(\ten)^2r}(i\,\epb_jC^{jk}\ggt\dot\T_k)-\frac{\dot r}{r}F.
\ee
Next a choice of $G=-\frac32F$ makes sum of the first three transformations as
\bea
{\D_{123}} L^{tot}&=&\pa_\tau\left(i\,r\,C^{jk}\epb_j\ggt\T_k+F(q^2+(\ten)^2r^2)-2G(\ten)^2r^2\right)+\frac{-\8\chi}{4(\ten)^2r}(i\,\epb_jC^{jk}\ggt\dot\T_k).
\nn\\
\label{del123L}\eea
Finally the last term of \bref{del123L} can be cancelled with the first term
of \bref{del4L} by a choice of $\8\rho_k$, using \bref{BBCC} and $k=\sqrt2 T$,
\bea
\8\rho_1=-\8\rho_2=\frac{1}{4kr}(\slP-kr\ggt)
\left((\B+1)\ep_1+(\B-1)\ep_2\right),
\eea
where we have used $\8\chi=-\frac14(\slP-kr\ggt)^2.$
Since the $\8\rho_k$'s are verifying  $C^{jk}\8\rho_k=0$
the second and third term of \bref{del4L} vanish as well. 

It completes a proof that the lagrangian \bref{Lagtotsuper} is invariant
under the four combined transformations.
The third and the forth transformations are essentially the diffeomorphism and 
the kappa transformations. The sum of first and second transformations
is the modified supersymmetry transformation generated by ${Q^*}^j$ 
in the hamiltonian formalism given in \bref{Qstar}.

\end{document}